\documentclass[11pt]{article}
\usepackage{longtable,supertabular}
\usepackage{amsmath}
\usepackage{amssymb}
\usepackage{latexsym}
\usepackage{cite}

\title{\bf On the Hull-Variation Problem of Equivalent Linear Codes}
\author{Hao Chen
  \thanks{Hao Chen is with the College of Information Science and Technology/Cyber Security, Jinan University, Guangzhou, Guangdong Province, 510632, China, haochen@jnu.edu.cn. The research of Hao Chen was supported by NSFC Grant 62032009.}}

\begin{document}

\maketitle
\begin{abstract}
The intersection ${\bf C}\bigcap {\bf C}^{\perp}$ (${\bf C}\bigcap {\bf C}^{\perp_h}$) of a linear code ${\bf C}$ and its Euclidean dual ${\bf C}^{\perp}$ (Hermitian dual ${\bf C}^{\perp_h}$) is called the Euclidean (Hermitian) hull of this code.  It is natural to consider the hull-variation problem when a linear code ${\bf C}$ is transformed to an equivalent code ${\bf v} \cdot {\bf C}$. In this paper we introduce the maximal hull dimension as an invariant of a linear code with respect to the equivalent transformations.  Then some basic properties of the maximal hull dimension are studied. We prove that for a nonnegative integer $h$ satisfying $0 \leq h \leq n-1$, a linear $[2n, n]_q$ self-dual code is equivalent to a linear $h$-dimension hull code. On the opposite direction we prove that a linear LCD code over ${\bf F}_{2^s}$ satisfying $d\geq 2$ and $d^{\perp} \geq 2$ is equivalent to a linear one-dimension hull code under a weak condition. Several new families of LCD negacyclic codes and LCD BCH codes over ${\bf F}_3$ are also constructed. Our method can be applied to the generalized Reed-Solomon codes and the generalized twisted Reed-Solomon codes to construct arbitrary dimension hull MDS codes. Some new entanglement-assisted quantum error-correction (EAQEC) codes including MDS and almost MDS EAQEC codes are constructed. Many EAQEC codes over small fields are constructed from optimal Hermitian self-dual codes.\\

\end{abstract}

\section{Introduction}

The Hamming weight $wt_H({\bf a})$ of a vector ${\bf a} \in {\bf F}_q^n$ is the number of non-zero coordinate positions. The Hamming distance $d_H({\bf a}, {\bf b})=wt_H({\bf a}-{\bf b})$ between two vectors ${\bf a}$ and ${\bf b}$ is the Hamming weight of ${\bf a}-{\bf b}$, that is, the number of coordinate positions where ${\bf a}$ and ${\bf b}$ are different. The Hamming distance of a code ${\bf C} \subset {\bf F}_q^n$, $$d_H({\bf C})=\min_{{\bf a} \neq {\bf b}} \{d_H({\bf a}, {\bf b}): {\bf a} \in {\bf C}, {\bf b} \in {\bf C} \},$$  is the minimum of Hamming distances $d_H({\bf a}, {\bf b})$ between any two different codewords in ${\bf C}$. The shortening code of ${\bf C}$ at the $i$-th coordinate position is the subcode of ${\bf C}$ consisting of codewords in ${\bf C}$ whose $i$-th coordinate is zero. The minimum Hamming distance of  a linear code is its minimum Hamming weight.  For a linear $[n, k, d_H]_q$ code, the Singleton bound asserts $d_H \leq n-k+1$. When the equality holds, this code is an MDS code. The main conjecture of MDS codes claims that the length of an MDS code over ${\bf F}_q$ is at most $q+1$, except some trivial exceptional cases. In \cite{Ball} the main conjecture of MDS codes was proved for codes over prime fields. We refer to \cite{Lint,HP} for the theory of Hamming error-correcting codes.\\

 Two codes ${\bf C}_1$ and ${\bf C}_2$ in ${\bf F}_q^n$ are equivalent if and only if ${\bf C}_2$ can be obtained from ${\bf C}_1$ by a permutation of coordinates and the multiplication of a Hamming weight $n$ vector ${\bf v}=(v_1, v_2, \ldots, v_n) \in {\bf F}_q^n$ on coordinates, where $v_i \neq 0$ for $i=1, \ldots, n$. That is $${\bf C}_2=\{{\bf c}=(c_1, \ldots c_n): (c_1, \ldots, c_n)=(v_1x_1, \ldots, v_nx_n), {\bf x} \in Perm({\bf C}_1)\},$$ where $Perm({\bf C}_1)$ is the code obtained from ${\bf C}_1$ by a coordinate permutation. Equivalent codes have the same parameters.\\

The Euclidean inner product on ${\bf F}_q^n$ is defined by $$<{\bf x}, {\bf y}>=\Sigma_{i=1}^n x_iy_i,$$ where ${\bf x}=(x_1, \ldots, x_n)$ and ${\bf y}=(y_1, \ldots, y_n)$. The Euclidean dual of a linear code ${\bf C}\subset {\bf F}_q^n$ is $${\bf C}^{\perp}=\{{\bf c} \in {\bf F}_q^n: <{\bf c}, {\bf y}>=0, \forall {\bf y} \in {\bf C}\}.$$ The Hermitian inner product on ${\bf F}_{q^2}^n$ is defined by  $$<{\bf x}, {\bf y}>_h=\Sigma_{i=1}^n x_iy_i^q,$$ where ${\bf x}=(x_1, \ldots, x_n)$ and ${\bf y}=(y_1, \ldots, y_n)$ are two vectors in ${\bf F}_{q^2}^n$. The Hermitian dual of a linear code ${\bf C} \subset {\bf F}_{q^2}^n$ is $${\bf C}^{\perp_h}=\{{\bf c} \in {\bf F}_{q^2}^n: <{\bf c}, {\bf y}>_h=0, \forall {\bf y} \in {\bf C}\}.$$  It is clear ${\bf C}^{\perp_h}=({\bf C}^{\perp})^q$, where $${\bf C}^q=\{(c_1^q, \ldots, c_n^q): (c_1, \ldots, c_n) \in {\bf C}\}.$$ The minimum distance of the Euclidean dual is called the dual distance and is denoted by $d^{\perp}$. The minimum distance of the Hermitian dual is the same as $d^{\perp}$. A linear code ${\bf C} \subset {\bf F}_q^n$ is linear complementary dual (LCD) code, if ${\bf C} \bigcap {\bf C}^{\perp}=\{{\bf 0}\}$,  self dual if ${\bf C}={\bf C}^{\perp}$,  and self-orthogonal if ${\bf C} \subset {\bf C}^{\perp}$. Similarly a linear code ${\bf C} \subset {\bf F}_{q^2}^n$ is called linear complementary Hermitian dual (Hermitian LCD) if ${\bf C} \bigcap {\bf C}^{\perp_h}=\{{\bf 0}\}$, Hermitian self-dual if ${\bf C}={\bf C}^{\perp_h}$ and Hermitian self-orthogonal if ${\bf C} \subset {\bf C}^{\perp_h}$. In general the linear code ${\bf C} \bigcap {\bf C}^{\perp}$ is called the Euclidean hull of the linear code ${\bf C} \subset {\bf F}_q$. The intersection ${\bf C} \bigcap {\bf C}^{\perp_h}$ is called the Hermitian hull of the linear code ${\bf C} \subset {\bf F}_{q^2}^n$.  It is obvious that an LCD code ${\bf C} \subset {\bf F}_q$ leads to a Hermitian LCD code ${\bf F}_{q^2} \otimes {\bf C}$.\\

We refer to \cite{CPS,CS}  for earlier results on self-dual codes over small fields and \cite{HP} Chapter 9 and \cite{Rains} for the theory of small length Euclidean and Hermitian self-dual code classification.  There have been a long-time  interest in coding theory to construct self-dual, Hermitian self-dual MDS codes, for example, see \cite{GG08,Gulliver,JX17,JK19,HY20,Sok,Sok2,Niu,ZhangFeng,GuoLi}. Obviously equivalent codes have different dual codes as follows. Let ${\bf v}$ be a Hamming weight $n$ vector ${\bf v}=(v_1, v_2, \ldots, v_n) \in {\bf F}_q^n$, set ${\bf v} \cdot {\bf C}=\{(v_1c_1, \ldots, v_nc_n): \forall {\bf c}=(c_1, \ldots, c_n) \in {\bf C}\}$. Then $$({\bf v} \cdot {\bf C})^{\perp}={\bf v}^{-1} \cdot {\bf C}^{\perp},$$ where ${\bf v}^{-1}=(v_1^{-1}, \ldots, v_n^{-1}).$ For a linear code ${\bf C} \subset {\bf F}_{q^2}^n$ and a Hamming weight $n$ vector ${\bf v} \in {\bf F}_{q^2}^n$, then $$({\bf v} \cdot {\bf C})^{\perp_h}={\bf v}^{-q} \cdot {\bf C}^{\perp_h},$$ where ${\bf v}^{-q}=(v_1^{-q}, \ldots, v_n^{-q})$. Notice that weight distributions and generalized Hamming weights, see \cite{Wei},  of two equivalent linear codes are the same, dual codes are not the same. Hence the motivation of this paper is to consider the following natural hull-variation problem.\\

{\bf Hull-variation problem.} {\em When a linear code ${\bf C}$ is transformed to an equivalent linear code ${\bf v} \cdot {\bf C}$, how its Euclidean or Hermitian hull is changed? }\\

The first motivation to consider the hull-variation problem is Sendrier's algorithm in \cite{Sen1} to determine all permutation-equivalences between two linear codes. From our main result Theorem 3.1 and Theorem 4.1, it is always possible to get an equivalent one-dimension hull linear code from a $h\geq 1$ dimension hull linear code, and it only needs a weak condition to get an equivalent one-dimension hull linear code from an LCD code over ${\bf F}_{2^s}$. Thus it seems that our results can help to determine the self permutation-equivalences of a linear code.\\

The second motivation to consider the hull-variation problem is from the construction of entanglement-assisted quantum codes. Quantum error correction codes (QECCs) are necessary for quantum information processing and quantum computation. The quantum error correction using classical error-correcting codes was proposed in \cite{Shor,Steane} and then the stabilizer Calderbank-Shor-Steane (CSS) construction of QECC was proposed in \cite{CShor,Steane1}.  We refer to \cite{HChen1,HChen2,HChen3,AKS,KS,JLLX,JX,Ball1} for constructions of QECCs from classical codes. In \cite{Brun,Fattal} entanglement-assisted quantum error correction (EAQEC) codes were proposed and studied. It was proved that EAQEC codes can be constructed from the stabilizer CSS construction. Comparing to an QECC, an EAQEC code has one more parameter $c$ measuring the consumption of $c$ pre-shared copies of maximally  entangled states. An EAQEC $[[n, k-h, d, n-k-h]]_q$ code can be obtained from a linear $[n, k, d]_q$ code with the $h$-dimension Euclidean hull. Similarly from a linear $[n, k, d]_{q^2}$ code ${\bf C} \subset {\bf F}_{q^2}^n$ with the $h$-dimension Hermitian hull, an EAQEC $[[n, k-h, d, n-k-h]]_q$ code can be constructed, see \cite{Brun}. The quantum Singleton bound asserts $$2d+k\leq n+c+2$$ for an EAQEC $[[n, k, d, c]]_q$ code when $d\leq \frac{n+2}{2}$, see \cite{Brun,GHW}.  In the above construction of EAQEC codes, the dimension $h$ of the hull of a linear code is a key parameter to determine the dimension and the consumption parameter of an EAQEC code, since the other parameters of codes are preserved when a linear code is transformed to an equivalent linear code. This is an important motivation to construct equivalent linear codes with various hull dimensions, see \cite{LCC,GYHZ,FFLZ20,CZJL,MCao}. An EAQEC code attaining the quantum Singleton bound is called an MDS EAQEC code. The construction of MDS EAQEC code with large ranges of four parameters has been addressed in \cite{LCC,GYHZ,Sok,Sok1,Wu}. An EAQEC code satisfying $2d+k=n+c$ and $d \leq \frac{n+2}{2}$ is called almost MDS. Some almost MDS EAQEC codes have been given in \cite{Pellikaan}. In our paper \cite{HChen22} it was proved that any QECC constructed from a Hermitian self-orthogonal code of the dimension $k$ can be transformed to $k$ different EAQEC codes with the same $n, d$ and $k$ different $c$ parameters.\\

LCD codes were introduced in 1964 by Massey in \cite{Massey} using the name reversible codes, and in his 1992 paper \cite{Massey1} using the name linear complementary dual codes. The existence of an asymptotically good sequence of LCD codes was proved in \cite{Massey1}. Based on the fundamental result proved in \cite{Carlet2}, it seems reasonable to ask if  one-dimension hull is a strong constraint on a linear code. It is also natural to ask what parameter $[n, k, d]_q$ can be attainable by a linear code with $h$-dimension hull for large or small fixed $h$. For a fixed positive integer $h$, is there an asymptotically good sequence of linear codes with the $h$-dimension hulls?  On the opposite direction we can ask for a linear Euclidean self-dual $[2n, n]_q$ code or a linear Hermitian self-dual $[2n, n]_{q^2}$ code, is there an equivalent $h<n$ dimension hull code?\\

LCD codes have applications in data storage, communication and cryptography, see \cite{Massey,Bringer,Carlet0}.  Cyclic LCD codes were studied in a paper \cite{YM} of Yang and Massey published in 1994. Sendrier proved that LCD codes meets the Gilbert-Varshamov bound in 2004. The application of LCD codes in resisting the side-channel attack proposed in \cite{Bringer,Carlet0} have stimulated a lot of constructions of LCD codes from cyclic codes and constacyclic codes, see \cite{LDL,LLDL,WY}, algebraic geometry codes, see \cite{MTQ}, Reed-Solomon codes, twisted Reed-Solomon codes, see \cite{J2017,LCC,GYHZ,ChenLiu,BJ,WHL} and J-affine variety codes, see \cite{GGHR}. LCD codes were counted in \cite{Carlet1,WY} and $\sigma$-LCD codes were introduced in \cite{Carlet}. Linear complementary pairs from constacyclic codes were studied in \cite{Carlet18}. Finally in \cite{Carlet2} it was proved that an arbitrary linear $[n, k, d]_q$ code over a finite field ${\bf F}_q$ satisfying $q>3$ is equivalent to an LCD code, and an arbitrary linear $[n, k, d]_{q^2}$
code over a finite filed ${\bf F}_{q^2}$, $q>2$, is equivalent to a Hermitian LCD code. A new approach of constructing LCD codes by extending linear codes was also presented in \cite{Carlet2}. This is quite different to the self-dual codes, where the fundamental Gleason theorem is a strong restriction on weight distributions of self-dual codes, see \cite{HP} Chapter 9 and \cite{Rains}. It is still interesting to construct new LCD binary and ternary codes after the publication of the paper \cite{Carlet2}. The above result in \cite{Carlet2} means that the hull-decreasing variation to zero has no obstruction, which can be considered as a special case of the hull-variation problem. The hull-variation problem of equivalent linear codes proposed in this paper is motivated from the paper \cite{Carlet2}. Moreover the hull-increasing variation of equivalent linear codes is considered in this paper. There were several recent papers \cite{LCC,GYHZ,Pellikaan,Sok,Sok1} about the construction of MDS linear codes with $h$-dimension hulls and their applications in EAQEC codes. Cyclic codes with one-dimension hulls or $h$-dimension hulls were studied in \cite{LZ,GLMQ,QianM}. Some MDS codes with arbitrary dimension hulls were constructed in \cite{Sok1,LCC,FFLZ20,HY20,Wu,SM22} for some restricted lengths.\\

In this paper we propose to study the hull-variation problem and introduce the maximal Euclidean (or Hermitian) hull dimension as an invariant of a linear code with respect to equivalent transformations. Some basic properties of maximal Euclidean (Hermitian) hull dimensions are presented. We prove that for a nonnegative integer $h$ satisfying $0 \leq h \leq n-1$, a linear $[2n, n]_q$ self-dual code is equivalent to a linear $h$-dimension hull code. On the opposite direction we prove that a linear LCD code over ${\bf F}_{2^s}$ satisfying $d\geq 2$ and $d^{\perp} \geq 2$ is equivalent to a linear one-dimension hull code under a weak condition. Many LCD codes satisfy this condition. Then it follows that for any fixed nonnegative integer $h$, there is an asymptotically good sequence of linear codes over ${\bf F}_q$ with the $h$-dimension hulls, when $q \geq 64$ and $q \neq 125$, whose asymptotically parameters exceeds the Gilbert-Varshomov bound. Our result directly implies the existence of many new entanglement-assisted quantum error correction codes with various parameters, some new MDS and new almost MDS entanglement-assisted quantum error correction codes. Several families of LCD negacyclic codes over ${\bf F}_q$ and several new families of LCD cyclic codes over ${\bf F}_q$, where $q$ is odd and in particular $q=3$, are constructed. Finally we show that our method can be used to construct arbitrary dimension hull MDS codes from the generalized Reed-Solomon codes and the generalized twisted Reed-Solomon codes over ${\bf F}_{2^s}$. Therefore we show that the hull-decreasing variation from self-dual codes has no obstruction and the hull-increasing variation from the dimension zero to the dimension one has a small obstruction. We also show that hull-increasing variations of Reed-Solomon codes and some twisted Reed-Solomon codes over ${\bf F}_{2^s}$ have no obstruction. It is an interesting problem to find the exact obstruction of the hull-increasing variation of equivalent linear codes.\\

\section{Maximal hull dimensions}

From the view of the hull-variation problem of equivalent linear codes, we can define the following maximal hull equivalent code.\\

{\bf Definition 2.1.} {\em Let ${\bf C} \subset {\bf F}_q^n$ (${\bf C} \subset {\bf F}_{q^2}^n$) be a linear $[n, k]_q$ ($[n, k]_{q^2}$) code, let ${\bf v}$ be a Hamming weight $n$ vector ${\bf v} \in {\bf F}_q^n$ (${\bf v} \in {\bf F}_{q^2}^n$) such that $\dim_{{\bf F}_q}(({\bf v} \cdot {\bf C}) \bigcap ({\bf v} \cdot {\bf C})^{\perp})$ ($\dim_{{\bf F}_{q^2}}(({\bf v} \cdot {\bf C}) \bigcap ({\bf v} \cdot {\bf C})^{\perp_h})$) is maximal. Then we call ${\bf v} \cdot {\bf C}$ a maximal Euclidean (Hermitian) hull equivalent code and the dimension $\dim_{{\bf F}_q}(({\bf v} \cdot {\bf C}) \bigcap ({\bf v} \cdot {\bf V})^{\perp})$ ($\dim_{{\bf F}_{q^2}}(({\bf v} \cdot {\bf C}) \bigcap ({\bf v} \cdot {\bf C})^{\perp_h})$) the maximal Euclidean (Hermitian) hull dimension $\max_{Ehull}({\bf C})$ ($\max_{Hhull}({\bf C})$).}\\

It is clear that the maximal Euclidean hull dimension (or the maximal Hermitian hull dimension) is an invariant of a linear code in ${\bf F}_q^n$ (or in ${\bf F}_{q^2}^n$) with respect to the equivalent transformations. When the code is equivalent to a self-orthogonal code, its maximal hull dimension is its own dimension. When ${\bf C} \subset {\bf F}_q^{2n}$ is an even length linear $[2n, n]_q$ code with the weight distribution $\{A_0, \ldots, A_{2n}\}$ not satisfying the MacWilliams equation of a self-dual code, see \cite{HP}, Chapter 7, (7.3),  $$\Sigma_{j=0}^{2n-v} \displaystyle{2n-j \choose v} A_j=q^{n-v}\Sigma_{j=0}^v \displaystyle{2n-j \choose 2n-v} A_j,$$ for $0 \leq v \leq 2n$, where $A_i$ is the number of weight $i$ codewords in ${\bf C}$, it is obvious that the maximal Euclidean hull dimension can not be $n$, since ${\bf C}$ can not be equivalent to a self-dual code. From this invariant we have the following natural question.\\

When the maximal Euclidean (or Hermitian) hull dimension of a Euclidean (or Hermitian) LCD code is $0$, we call it a rigid LCD code.\\

{\bf Maximal hull-variation problem.} {\em For a linear code ${\bf C} \subset {\bf F}_q^n$ (or a linear code ${\bf C} \subset {\bf F}_{q^2}^n$), to find the maximal Euclidean hull dimension $\max_{Ehull}({\bf C})$ (or the maximal Hermitian hull dimension $\max_{Hhull}({\bf C})$).}\\

The following two propositions are obvious.\\

{\bf Proposition 2.1.} {\em Let ${\bf C} \subset {\bf F}_q^n$ be a linear code then $\max_{Ehull}({\bf C}) \leq \dim({\bf C})$. Let ${\bf C} \subset {\bf F}_{q^2}^n$ be a linear code then $\max_{Hhull}({\bf C}) \leq \dim({\bf C})$. We have $\max_{Ehull}({\bf C})=\max_{Ehull}({\bf C}^{\perp})$ and $\max_{Hhull}({\bf C})=\max_{Hhull}({\bf C}^{\perp_h})$. When ${\bf C}$ is self-orthogonal then $\max_{Ehull}({\bf C})=\dim({\bf C})$ and $\max_{Hhull}({\bf C})=\dim({\bf C})$.}\\

{\bf Proposition 2.2.} {\em Let ${\bf C} \subset {\bf F}_q^n$ be a linear code of the dimension $1$. Then $\max_{Ehull}({\bf C})=1$ when $d_H({\bf C}) \geq 2$ and $\max_{Ehull}({\bf C})=0$ when $d_H({\bf C})=1$. Let ${\bf C} \subset {\bf F}_{q^2}^n$ be a linear code of dimension $1$. Then $\max_{Hhull}({\bf C})=1$ when $d_H({\bf C}) \geq 2$ and $\max_{Hhull}({\bf C})=0$ when $d_H({\bf C})=1$.}\\

Then a dimension one linear LCD code is rigid if and only if its minimum Hamming distance is $1$.\\

Let ${\bf C}_1 \star {\bf C}_2$ be the componentwise product of two linear codes ${\bf C}_1$ and ${\bf C}_2$ in ${\bf F}_q^n$. It is the linear span in ${\bf F}_q^n$ of all vectors of the form $(c_1 c_1', \ldots, c_n c_n')$ where ${\bf c}=(c_1, \ldots, c_n) \in {\bf C}_1$ and ${\bf c}'=(c_1', \ldots, c_n') \in {\bf C}_2$. This kind of linear code was studied in \cite{Ran} and played an important role in secure multiparty computation, see \cite{CCCX}. When ${\bf C}_1={\bf C}_2$, this is called the Schur square of this linear code. It is clear that the dimension of the Schur square of a linear code is an invariant with respect to code equivalence transformations. The dimension of the Schur square of a Reed-Solomon code $RS(n, k)$ is $2k-1$, see \cite{BPR} and references there in.\\

{\bf Theorem 2.1.} {\em Let $q=2^s$ be an even prime power and ${\bf C} \subset {\bf F}_q^n$ be a linear $[n, k]_q$ code with its dimension $k \leq \frac{n}{2}$. Let ${\bf C}_1 \subset {\bf C}$ be a linear subcode of the dimension $k_1$. If there is a Hamming weight $n$ vector in the dual of the Schur square ${\bf C}_1 \star {\bf C}_1$ of this linear subcode, then $\max_{Ehull}({\bf C}) \geq k_1$. Let $q$ be an arbitrary prime power and ${\bf C} \subset {\bf F}_{q^2}^n$ be a linear $[n, k]_{q^2}$ code.  Let ${\bf C}_1 \subset {\bf C}$ be a linear subcode of the dimension $k_1$. If there is a Hamming weight $n$ vector in the subfield subcode $({\bf C}_1 \star {\bf C}_1^q)^{\perp} \bigcap {\bf F}_q^n$ of the dual of the componentwise product of ${\bf C}_1$ and ${\bf C}_1^q$, then $\max_{Hhull}({\bf C}) \geq k_1$.}\\

{\bf Proof.} We need to find $n$ nonzero elements $v_1, \ldots, v_n$ such that $c_1 c_1' v_1^2+\cdots+c_n c_n' v_n^2=0$ for any two codewords ${\bf c}=(c_1, \ldots, c_n) \in {\bf C}$ and ${\bf c}'=(c_1, \ldots, c_n') \in {\bf C}_1$. When $q=2^s$ is an even prime power, then $v_1^2, \ldots, v_n^2$ can be arbitrary nonzero elements in ${\bf F}_{2^s}$, since each element in ${\bf F}_{2^s}$ is a square. The conclusion follows directly. The conclusion for the maximal Hermitian hull dimension can be proved similarly.\\

Let ${\bf C}=\{({\bf x}, {\bf x}): {\bf x} \in {\bf F}_q^n\}$ be a linear $[2n, n]_q$ code, where $q$ is an even prime power. It is clear that ${\bf C} \star {\bf C}={\bf C}$, whose dual contains a Hamming weight $n$ codeword since $-1=1$ is a square in ${\bf F}_q$, since $q$ is even. Then $\max_{EHull}({\bf C})=n$. Similarly for the above code over ${\bf F}_{q^2}$. ${\bf C} \star {\bf C}^q={\bf C}$, then $\max_{Hhull}({\bf C})=n$. When ${\bf C}$ is a Reed-Solomon $[n, k]_{2^s}$ code with the dimension $k <\frac{n}{2}$, then ${\bf C} \star {\bf C}$ is a Reed-Solomon $[n, 2k-1]_{2^s}$ code. From Theorem 2.1 $\max_{Ehull}({\bf C})=k$. This can be generalized to algebraic-geometric codes as follows.\\

Let ${\bf X}$ be an absolutely irreducible, smooth genus $g$ curve defined over ${\bf F}_q$. Let ${\bf P}=\{P_1,\ldots,P_n\}$ be $n$ distinct rational points of ${\bf X}$ over ${\bf F}_q$. Let ${\bf G}$ be a rational effective divisor over ${\bf F}_q$ of degree $\deg({\bf G})$ satisfying $2g-2 <\deg({\bf G})<n$ and $$support({\bf G}) \bigcap {\bf P}=\emptyset.$$ Let ${\bf L}({\bf G})$ be the function space associated with the divisor ${\bf G}$. The algebraic-geometric code associated with ${\bf G}$, ${\bf P}=\{P_1,\ldots,P_n\}$ is defined by $${\bf C}({\bf P}, {\bf G}, {\bf X})={\bf C}(P_1,\ldots,P_n, {\bf G}, {\bf X})=\{(f(P_1),\ldots,f(P_n)): f \in {\bf L}({\bf G})\}.$$ The dimension of this code is $$k=\deg({\bf G})-g+1$$ follows from the Riemann-Roch Theorem. The minimum Hamming distance is $$d_H \geq n-\deg({\bf G}).$$ Let ${\bf G}=\Sigma m_i Q_i$, $m_i \geq 0$ be an effective divisor, if ${\bf G}_1=\Sigma m_i' Q_i$ is another effective divisor satisfying $m_i' \leq m_i$ we say ${\bf G}_1 \leq {\bf G}$. It is obvious that $${\bf C}(P_1,\ldots,P_n, {\bf G}_1, {\bf X})  \subset {\bf C}(P_1,\ldots,P_n, {\bf G}, {\bf X})$$ is a linear subcode. The Reed-Solomon codes are just the algebraic-geometric codes over the genus $0$ curve. One achievement of the theory of algebraic-geometric codes is the sequence of algebraic-geometric codes over ${\bf F}_{q^2}$ satisfying the Tsfasman-Vl\'{a}dut-Zink bound $$R+\delta \geq 1-\frac{1}{q-1},$$ which is exceeding the Gilbert-Varshamov bound when $q \geq 7$.  We refer to \cite{HP} Chapter 13 and \cite{Lint} Chapter 10 for the detail.\\

{\bf Theorem 2.2.} {\em Let ${\bf C}(P_1, \ldots, P_n, {\bf G}, {\bf X})$ be an algebraic-geometric code defined over ${\bf F}_{2^s}$. Let ${\bf G}_1 \leq {\bf G}$ be an effective  divisor satisfying $$2g-2<\deg({\bf G}_1) \leq g-1+\lfloor\frac{n}{2} \rfloor.$$ Then this algebraic-geometric code ${\bf C}(P_1, \ldots, P_n, {\bf G}, {\bf X})$ has its maximal Euclidean hull dimension at least $k_1=\dim({\bf C}(P_1, \ldots, P_n, {\bf G}_1, {\bf X})=\deg({\bf G}_1)-g+1$.}\\

{\bf Proof.} Its Schur square is in the algebraic-geometric code ${\bf C}(P_1, \ldots, P_n, \\2{\bf G}_1, {\bf X})$. The conclusion follows immediately.\\

\section{The hull-decreasing variation from self-dual or Hermitian self-dual codes}

Let ${\bf A}=(a_{ij})_{1 \leq i, j \leq n}$ be an $n \times n$ matrix with entries $a_{ij} \in {\bf F}_{q^2}$, then the conjugate matrix is ${\bf \bar{A}}=(a_{ij}^q)_{1 \leq i, j \leq n}$. The following result is well-known, see \cite{Carlet2}.\\

{\bf Proposition 3.1.} {\em Let $({\bf I}_n, {\bf P})$ be a generator matrix of a linear self-dual $[2n, n]_q$ code ${\bf C} \subset {\bf F}_q^n$. Then ${\bf P} \cdot {\bf P}^{\tau}=-{\bf I}_n$. Therefore ${\bf P}$ is a non-singular $n \times n$ matrix. Let $({\bf I}_n, {\bf A})$ be a generator matrix of a linear Hermitian self-dual $[2n, n]_{q^2}$ code ${\bf C} \subset {\bf F}_{q^2}^n$. Then ${\bf A} \cdot {\bf \bar{A}}^{\tau}=-{\bf I}_n$. Therefore ${\bf A}$ is a non-singular $n \times n$ matrix.}\\

The main result of this Section is as follows. \\

{\bf Theorem 3.1.} {\em Let $q \geq 3$ be a prime power. Let $n$ be a positive integer and $h$ be a nonnegative integer satisfying $0 \leq h <n$. Let ${\bf C} \subset {\bf F}_q^n$ be a linear self-dual $[2n, n]_q$ code. Then there is a Hamming weight $n$ vector ${\bf v} \in {\bf F}_q$ such that the linear code ${\bf v} \cdot {\bf C}$ has the $h$-dimension hull. Let ${\bf C} \subset {\bf F}_{q^2}^n$ be a linear Hermitian self-dual $[2n, n]_{q^2}$ code. Then there is a Hamming weight $n$ vector ${\bf v} \in {\bf F}_{q^2}$ such that the linear code ${\bf v} \cdot {\bf C}$ has the $h$-dimension Hermitian hull.}\\

{\bf Proof.} We choose a generator matrix $({\bf I}_n, {\bf P})$ of the linear self-dual $[2n, n]_q$ code ${\bf C} \subset {\bf F}_q^n$. Set ${\bf v}=(\lambda_1, \ldots, \lambda_{n-h}, 1, \ldots, 1)$, where $\lambda_1, \ldots, \lambda_{n-h}$ are $n-h$ nonzero elements in ${\bf F}_q$ such that all $\lambda_1^2, \ldots, \lambda_{n-h}^2$ are not $1$. Then a generator matrix of ${\bf v} \cdot {\bf C}$ is of the following form $({\bf D_{\lambda}}, {\bf P})$, where ${\bf D_{\lambda}}$ is a $n \times n$ diagonal non-singular matrix of the following form.\\

$$
\left(
\begin{array}{ccccccccccccc}
\lambda_1&0&0&\cdots&\cdots&\cdots&\cdots&0\\
0&\lambda_2&0&\cdots&\cdots&\cdots&\cdots&0\\
\cdots&\cdots&\cdots&\cdots&\cdots&\cdots&\cdots&\cdots\\
0&0&0&\cdots&\lambda_{n-h}&0&\cdots&0\\
0&0&0&\cdots&0&1&\cdots&0\\
0&0&0&\cdots&0&0&\cdots&1\\
\end{array}
\right)
$$
The dual is ${\bf v}^{-1} \cdot {\bf C}$ with the generator matrix $({\bf D_{\lambda}}^{-1}, {\bf P})$. It is clear  that the last $h$ rows of these two generator matrices are the same. Thus the dimension of the hull $({\bf v} \cdot {\bf C}) \bigcap ({\bf v} \cdot {\bf C})^{\perp}$ is at least $h$. Let ${\bf C}_1$ be the $n-h$ dimension subcode of ${\bf v} \cdot {\bf C}$ generated by the first $n-h$ rows of the above generator matrix. We now prove that the dimension of the hull is exactly $h$. Otherwise the natural mapping $$({\bf v} \cdot {\bf C}) \bigcap ({\bf v} \cdot {\bf C})^{\perp} \longrightarrow ({\bf v} \cdot {\bf C})/{\bf C}_1$$ is not injective and there is a nonzero ${\bf c} \in ({\bf v} \cdot {\bf C}) \bigcap ({\bf v} \cdot {\bf C})^{\perp}$ such that ${\bf c} \subset {\bf C}_1$. Then we have a nonzero vector ${\bf x} \in {\bf F}_q^n$ such that its last $h$ coordinates are zero and a nonzero vector in ${\bf y} \in {\bf F}_q^n$, satisfying ${\bf x} \cdot ({\bf D_{\lambda}}, {\bf P})={\bf y}({\bf D_{\lambda}}^{-1}, {\bf P})$. From the reading of the first $n$ coordinates of this codeword, the last $h$ coordinates of ${\bf y}$ are zero. Let ${\bf P}'$ be the $(n-h) \times n$ sub-matrix of ${\bf P}$ consisting of its first $n-h$ rows. This is a rank $n-h$ matrix since ${\bf P}$ is nonsingular. Then we have ${\bf x} \cdot  {\bf P}'={\bf y} \cdot {\bf P}'$ from the reading of the last $n$ coordinates of this codeword. Therefore ${\bf x}={\bf y}$ since ${\bf P}'$ is a full-rank matrix. From the reading of the first $n-h$ coordinates, we have $\lambda_i^2=1$. This is a contradiction. The conclusion about the Euclidean hull is proved. The conclusion about the Hermitian hull can be proved similarly.\\

The similar result was proved in Theorem 7 of \cite{LEGL22} from the physical motivation of the entanglement-assisted quantum code.\\

There is a length $2n$ self-dual MDS generalized Reed-Solomon code over ${\bf F}_{2^s}$ from the classical result in \cite{GG08,Gulliver}. Therefore from Theorem 3.1, for any even length $2n \leq 2^s$, there is an equivalent $h$-dimension hull $[2n, n]_{2^s}$ MDS code for a nonnegative integer $0 \leq h <n$. When $q$ is an odd prime power, self-dual MDS codes with the various lengths have been constructed, see \cite{JX17,ZhangFeng,Sok}. Therefore from Theorem 3.1, there are equivalent arbitrary dimension hull MDS codes for all these self-dual MDS codes presented in \cite{JX17,ZhangFeng,Sok} and references therein. Some of them are new comparing with the constructions in \cite{LCC,FFLZ20}. For Hermitian self-dual codes, only few small length Hermitian self-dual GRS codes have been constructed in \cite{Niu,GuoLi}. From Theorem 3.1 there are equivalent arbitrary dimension Hermitian hull GRS codes. On the other hand some self-dual MDS twisted Reed-Solomon codes were constructed in \cite{WHL}. From Theorem 3.1 arbitrary dimension hull generalized twisted Reed-Solomon codes can be constructed. These arbitrary dimension hull generalized twisted Reed-Solomon codes are new and can be compared with codes constructed in Subsection 6.2.\\

For asymptotically good linear $h$-dimension codes, we have the following result immediately from Theorem 3.1 and the main results in \cite{Bassa,Stich}.\\

{\bf Corollary 3.1.} {\em Let $q$ be a prime power satisfying $q \geq 64$ and $q \neq 125$, and $h$ be a fixed non-negative integer. Then there is an asymptotically good sequences of linear $h$-dimension hull codes exceeding the Gilbert-Varshamov bound.}\\

The above result is a generalization of previous results in \cite{Massey1,Carlet2}.\\

\section{The hull-increasing variation from LCD codes}

Set ${\bf v}=(\lambda, 1, \ldots, 1) \in {\bf F}_q^n$, $\lambda \in {\bf F}_q$ is a nonzero element. For a code ${\bf C} \subset {\bf F}_q^n$, we call the code ${\bf v} \cdot {\bf C}$ the $\lambda$-disturbing code of ${\bf C}$. It is clear that the dual of a $\lambda$-disturbing code of ${\bf C}$ is a $\lambda^{-1}$-disturbing code of the dual ${\bf C}^{\perp}$. The Hermitian dual of the $\lambda$-disturbing code of ${\bf C}$ is the $\lambda^{-q}$-disturbing code of the Hermitian dual ${\bf C}^{\perp_h}$.\\

{\bf Theorem 4.1.} {\em Let $q$ be an even prime power satisfying $q \geq 4$. Let ${\bf C} \subset {\bf F}_q^n$ be a linear LCD $[n, k, d]_q$ code with the minimum Hamming distance $d\geq 2$ and its dual distance $d^{\perp} \geq 2$. Suppose that the shortening codes of both ${\bf C}$ and ${\bf C}^{\perp}$ at one coordinate position are LCD codes. Then there is a $\lambda$-disturbing code of  ${\bf C}$ which has the one-dimension Euclidean hull.}\\

{\bf Proof.} Let ${\bf C} \subset {\bf F}_q^n$ be an LCD code of the dimension $k$. First of all we assume that the generator matrices of ${\bf C}$ and ${\bf C}^{\perp}$ are of the following forms,

$$
\left(
\begin{array}{cccccc}
1&{\bf a}\\
0&{\bf G}_1\\
\end{array}
\right)
$$
where ${\bf a} \in {\bf F}_q^{n-1}$ and ${\bf G}_1$ is a $(k-1) \times (n-1)$ matrix over ${\bf F}_q$,

$$
\left(
\begin{array}{cccccc}
c&{\bf b}\\
0&{\bf G}_2\\
\end{array}
\right)
$$
where $c$ is a constant, ${\bf b} \in {\bf F}_q^{n-1}$ and ${\bf G}_2$ is a $(n-k-1) \times (n-1)$ matrix over ${\bf F}_q$. Let ${\bf A}$ be the first row of the above generator matrix of ${\bf C}$ and ${\bf B}$ be the first row of the above generator matrix of ${\bf C}^{\perp}$. Let ${\bf a} \in {\bf F}_q^{n-1}$ be the vector of the last $n-1$ coordinates of ${\bf A}$ and ${\bf b} \in {\bf F}_q^{n-1}$ be the vector of the last $n-1$ coordinates of ${\bf B}$.  There is a linear subcode ${\bf C}_2 \subset {\bf C}$ of the dimension $k-1$ generated by the $(k-1) \times n$ matrix $({\bf 0}, {\bf G}_1)$,  and a linear subcode ${\bf C}_3 \subset {\bf C}^{\perp}$ of the dimension $n-k-1$ generated by the $(n-k-1) \times n$ matrix $({\bf 0}, {\bf G}_2)$.  These two codes are the shortening codes of ${\bf C}$ and ${\bf C}^{\perp}$ at the first coordinate position. It is obvious that we have ${\bf C}_2 \subset {\bf C}$ and ${\bf C}_3 \subset {\bf C}^{\perp}$. Thus ${\bf C}_2 \bigcap {\bf C}_3 =\{{\bf 0}\}$ since ${\bf C}$ is an LCD code.\\

The constant $c$ is not zero since the the minimum Hamming distance of ${\bf C}$ is at least $2$. It is easy to verify that the dual code of ${\bf C}_2$ is the code generated by ${\bf b}$ and ${\bf C}_3$, and the dual code of ${\bf C}_3$ is generated by ${\bf a}$ and ${\bf C}_2$. Since we assume that both ${\bf C}_2$ and ${\bf C}_3$ are LCD codes, then $n-1$ vectors,  ${\bf a}$, $k-1$ rows of the matrix ${\bf G}_1$, and $n-k-1$ rows of ${\bf G}_2$ in ${\bf F}_q^{n-1}$ are linearly independent, since ${\bf C} \bigcap {\bf C}_3={\bf 0}$. Similarly $n-1$ vectors, $k-1$ rows of the matrix ${\bf G}_1$, ${\bf b}$ and the $n-k-1$ rows of ${\bf G}_2$ in ${\bf F}_q^{n-1}$ are linearly independent, since ${\bf C}_2 \bigcap {\bf C}^{\perp}={\bf 0}$.\\

We will prove that there is a nonzero constant $\lambda \in {\bf F}_q$ such that the $\lambda$-disturbing of ${\bf C}$ is a linear one-dimension hull code. Set ${\bf C}_1={\bf v} \cdot {\bf C}$, then ${\bf C}_1^{\perp}={\bf v}^{-1} \cdot {\bf C}^{\perp}$ is a $\lambda^{-1}$-disturbing code of the dual ${\bf C}^{\perp}$.  It is clear that ${\bf C}_2$ is a subcode of ${\bf C}_1$ and ${\bf C}_3$ is a subcode of ${\bf C}_1^{\perp}$. To prove the conclusion of Theorem 4.1 we first prove the follow Lemma 4.1.\\

{\bf Lemma 4.1.} {\em For any nonzero constant $\lambda$, the dimension of the hull of ${\bf C}_1$ is at most $1$.}\\

{\bf Proof.} The natural linear mapping ${\bf C}_1 \bigcap {\bf C}_1^{\perp} \longrightarrow {\bf C}_1/{\bf C}_2 \oplus {\bf C}_1^{\perp}/{\bf C}_3$ is injective. Hence the dimension of the hull of ${\bf C}_1$ is at most $2$. On other hand if the dimension of the hull of ${\bf C}_1$ is $2$, the natural linear mapping ${\bf C}_1 \bigcap {\bf C}_1^{\perp} \longrightarrow {\bf C}_1/{\bf C}_2$ has a one-dimension kernel. Then there exists a nonzero codeword ${\bf x}$ of  ${\bf C}_2 \subset {\bf C}$ which is also in ${\bf C}_1^{\perp}$. Then ${\bf x}$ is in ${\bf C}^{\perp}$. This is a contradiction.\\

Now we need to prove that for a suitable $\lambda$, there is a nonzero vector in the intersection ${\bf C}_1 \bigcap {\bf C}_1^{\perp}$. The generator matrix of ${\bf C}_1$ is of the following form,

$$
\left(
\begin{array}{cccccc}
\lambda&{\bf a}\\
0&{\bf G}_1\\
\end{array}
\right)
$$
and the generator matrix of ${\bf C}_1^{\perp}$ is of the following form,
$$
\left(
\begin{array}{cccccc}
\lambda^{-1}c&{\bf b}\\
0&{\bf G}_2\\
\end{array}
\right)
$$
Let ${\bf A}_1$ be the first row of the above generator matrix of ${\bf C}_1$ and ${\bf B}_1$ be the first row of the above generator matrix of ${\bf C}_1^{\perp}$. Then we have $${\bf A}_1={\bf A}+(\lambda-1){\bf e},$$ and $${\bf B}_1={\bf B}+c(\frac{1}{\lambda}-1){\bf e},$$ where ${\bf e}=(1, 0, \ldots, 0) \in {\bf F}_q^n$. We need to find $u_1$ and $u_2$ in ${\bf F}_q$ satisfying $$u_1{\bf A}_1+{\bf c}_1=u_2{\bf B}_1+{\bf c}_2,$$ where ${\bf c}_1 \in {\bf C}_2$ and ${\bf c}_2 \in {\bf C}_2$ and at least one of $u_1$ and $u_2$ is not zero.  Here the constant $c$ is not zero, since $d\geq 2$ is assumed.\\

 The code ${\bf C}$ is an LCD code then ${\bf C} \oplus {\bf C}^{\perp}={\bf F}_q^n$. There exist $w_1$ and $w_2$ in ${\bf F}_q$, and ${\bf c}_1' \in {\bf C}_2$ and ${\bf c}_2' \in {\bf C}_3$ such that $${\bf e}=w_1{\bf A}+w_2{\bf B}+{\bf c}_1'+{\bf c}_2'.$$ Therefore we have $w_1+w_2c=1$. If $w_1$ or $w_2$ is zero, we show that the required nonzero codeword in ${\bf C}_1 \bigcap {\bf C}_1^{\perp}$ can be found as follows. If $w_2=0$, then $w_1$ is not $1$, otherwise the $n-1$ vectors in ${\bf F}_q^{n-1}$, ${\bf a}$, the $k-1$ rows of ${\bf G}_1$, and $n-k-1$ rows of ${\bf G}_2$ are linear dependent. This is a contradiction. Thus from the vector $(w_1-1, {\bf a}) \in {\bf F}_q^n$ with $w_1-1 \neq 0$ the required nonzero codeword in ${\bf C}_1 \bigcap {\bf C}_1^{\perp}$ can be constructed by setting $\lambda=w_1-1$. Similarly if $w_1=0$, then $w_2c \neq 1$, otherwise the $n-1$ vectors in ${\bf F}_q^{n-1}$, the $k-1$ rows of ${\bf G}_1$, ${\bf b}$ and $n-k-1$ rows of ${\bf G}_2$ are linearly dependent. This is a contradiction. The required nonzero codeword in ${\bf C}_1 \bigcap {\bf C}_1^{\perp}$ can be constructed by setting $\lambda^{-1}=w_2c-1$. Therefore from now on we assume $w_1\neq 0, 1$, and $w_2c \neq 0, 1$.\\

We need to find a $\lambda$ satisfying the following equation $$w_1(\lambda-1)+w_2c(\lambda^{-1}-1)=-1.$$ Set $w_2c=s$, then $s$ and $1-s$ is not zero. The above equation is equivalent to $$(s-1)\lambda^2=s.$$ Since $q$ is an even prime power, then each nonzero element in ${\bf F}_q$ is a square. The conclusion is proved.\\

Notice that the requirement $d^{\perp} \geq 2$ and $d \geq 2$ can be verified quickly. This is equivalent that in both the generator matrix and the parity check matrix of this code, there is no zero column.  Then from Theorem 4.1 some constructed BCH LCD codes and MDS LCD codes over the finite field ${\bf F}_q$ in \cite{LDL,LLDL}, where $q$ is even,  can be transformed to equivalent one-dimension hull codes by $\lambda$-disturbing codes.\\

\section{Ternary LCD cyclic and negacyclic codes}

After the fundamental results proved in \cite{Carlet2} it is still interesting to construct binary and ternary LCD codes. We refer to \cite{GGHR} on this topic. From the view of the hull-variation problem of cyclic codes we construct some ternary LCD cyclic and negacyclic codes from the previously constructed LCD cyclic codes in \cite{LLDL,LDL}. Though it was proved in \cite{Carlet2} that any linear $[n, k, d]_q$ code is equivalent to an LCD code when $q>3$, it is still interesting to see what parameters can be realized by binary and ternary LCD codes. It is also interesting to classify binary and ternary LCD BCH codes, LCD cyclic and LCD negacyclic codes.\\

 A code ${\bf C} \subset {\bf F}_q^n$ is called cyclic if $(c_0, c_1, \ldots, c_{n-1}) \in {\bf C}$, then $(c_{n-1}, c_0, \ldots, \\c_{n-2}) \in {\bf C}$. A codeword ${\bf c}$ in a cyclic code is identified with a polynomial ${\bf c}(x)=c_0+c_1x+\cdots+c_{n-1}x^{n-1}\in {\bf F}_q[x]/(x^n-1)$. Every cyclic code is a principal ideal in the ring ${\bf F}_q[x]/(x^n-1)$ and then generated by a factor of $x^n-1$. The code with the generator polynomial ${\bf g}(x)=g_0+g_1x+\cdots+g_{n-k}x^{n-k} \in {\bf F}_q[x]$ is denoted by  ${\bf C}_{{\bf g}}$. The dimension of the cyclic code ${\bf C}_{{\bf g}}$ generated by ${\bf g}(x)$ is $n-\deg({\bf g}(x))=k$. One generator matrix of ${\bf C}_{{\bf g}}$ is of the following form.\\

$$
\left(
\begin{array}{cccccccccccccccccccccc}
g_0&g_1&g_2&\cdots&\cdots&g_{n-k}&0&\cdots&\cdots&\cdots&\cdots&0\\
0&g_0&g_1&g_2&\cdots&\cdots&g_{n-k}&0&\cdots&\cdots&\cdots&0\\
\cdots&\cdots&\cdots&\cdots&\cdots&\cdots&\cdots&\cdots&\cdots&\cdots\cdots&\cdots\\
0&\cdots&\cdots&\cdots&0&g_0&g_1&g_2&0&0&\cdots&g_{n-k}\\
\end{array}
\right)
$$

The dual code of a cyclic code ${\bf C}_{{\bf g}}$ is a cyclic code with the generator polynomial ${\bf g}^{\perp}=\frac{x^k{\bf h}(x^{-1})}{{\bf h}(0)}$, where ${\bf h}(x)=\frac{x^n-1}{{\bf g}(x)}$. Therefore the root of ${\bf g}^{\perp}$ is of the form $\frac{1}{\beta}$ if $\beta$ is not a root of ${\bf g}(x)$, where $\beta$ is an $n$-th root of $1$ in some extension field of ${\bf F}_q$, see \cite{HP} Chapter 4.\\

Let ${\bf F}_q$ be a finite field and $\lambda \in {\bf F}_q$ be a nonzero constant. Then a linear code ${\bf C} \subset {\bf F}_q^n$ is called $\lambda$-cyclic if $(c_0, c_1, \ldots, c_{n-1}) \in {\bf C}$, then $(\lambda c_{n-1}, c_0, \ldots, \\c_{n-2}) \in {\bf C}$. A codeword ${\bf c}$  is identified with a polynomial ${\bf c}(x)=c_0+c_1x+\cdots+c_{n-1}x^{n-1}\in {\bf F}_q[x]/(x^n-\lambda)$. Every $\lambda$-cyclic code is generated by a factor of $x^n-\lambda$. The dimension of the $\lambda$-cyclic code ${\bf C}_{{\bf g}}$ generated by ${\bf g}(x)=g_0+g_1x+\cdots+g_{n-k}x^{n-k} \in {\bf F}_q[x]$ is $n-\deg({\bf g}(x))=k$. The dual of a $\lambda$-cyclic code is a $\lambda^{-1}$-cyclic code generated by ${\bf g}^{\perp}=\frac{x^k{\bf h}(x^{-1})}{{\bf h}(0)}$, where ${\bf h}(x)=\frac{x^n-\lambda}{{\bf g}(x)}$. It is well- known that if $\lambda^2 \neq 1$, then a $\lambda$-cyclic code is LCD. When $q$ is an odd prime power and $\lambda=-1$, a $(-1)$-cyclic code is called negacyclic.  Though many LCD cyclic BCH codes were constructed in \cite{LLDL,LDL} and it was proved that any linear $[n, k, d]_q$ code is equivalent to an LCD code when $q>3$. No previous known family of ternary negacyclic LCD codes has been constructed.\\

Let $q$ be a prime power, and $n$ be a positive integer satisfying $\gcd(n, q)=1$. Let $m$ be the smallest positive integer such that $q^m \equiv 1$ $mod$ $n$. Let $\alpha$ be a primitive element of the extension field ${\bf F}_{q^m}$. Then $\beta=\alpha^{\frac{q^m-1}{n}}$ is a primitive $n$-th root of unity. Let ${\bf m}_i(x)$ be the minimal polynomial of $\beta^i$ over ${\bf F}_q$. For a given designed distance $\delta \geq 2$, we set $${\bf g}_{q, n, \delta, b}(x)=lcm({\bf m}_b(x), {\bf m}_{b+1}(x), \ldots, {\bf m}_{b+\delta-2}(x)),$$ where $lcm$ is the least common multiple of polynomials in ${\bf F}_q[x]$. The cyclic code ${\bf C}_{q, n, \delta, b}$ generated by ${\bf g}_{q, n, b, \delta}(x)$ is called Bose-Chaudhuri-Hocquenghem (BCH) code introduced in \cite{BC1,BC2,Hoc}. It was proved in \cite{BC1,BC2,Hoc} the minimum Hamming distance is larger than or equal to the designed distance. The BCH codes have been studied for many years and served as basic examples of good linear error-correcting codes. Parameters of many LCD BCH codes have been determined in \cite{LLDL,LDL}.  From the following result, odd length cyclic codes over ${\bf F}_q$, where $q$ is an odd prime power, in Theorem 13 of \cite{LDL} can be transformed to negacyclic codes immediately.\\

{\bf Theorem 5.1.} {\em Let $q$ be an odd prime power and ${\bf C} \subset {\bf F}_q^n$ be a cyclic LCD code with the odd length $n$. Suppose that ${\bf C}$ is generated by a factor ${\bf g}(x)$ of $x^n-1$. Then we have a negacyclic LCD code ${\bf C}'$ generated  by $g(-x)$.}\\

{\bf Proof.} Since ${\bf g}(x)$ is a factor of $x^n-1$, then ${\bf g}(-x)$ is a factor of $(-x)^n-1=-(x^n+1)$. The dual of the negacyclic code ${\bf C}_{{\bf g}(-x)}$ is generated by ${\bf g}^{\perp}(-x)$. Let ${\bf v}=(1, -1, \ldots, -1, 1) \in {\bf F}_q^n$ be a Hamming weight $n$ vector. It is clear that ${\bf v}^{-1}={\bf v}$. Then the negacyclc code ${\bf C}_{{\bf g}(-x)}$ generated by $g(-x)$ is just ${\bf v} \cdot {\bf C}$ and its dual is ${\bf v}^{-1} \cdot {\bf C}^{\perp}={\bf v} \cdot {\bf C}^{\perp}$. Since ${\bf C}$ is an LCD code then ${\bf C} \bigcap {\bf C}^{\perp}=\{{\bf 0}\}$. Thus we have  $${\bf v} \cdot {\bf C} \bigcap {\bf v} \cdot {\bf C}^{\perp}={\bf v} \cdot {\bf C} \bigcap ({\bf v} \cdot {\bf C})^{\perp}=\{{\bf 0}\}.$$ The conclusion is proved.\\

From the following Corollaries several families of new ternary LCD negacyclic and LCD cyclic codes are constructed.\\

{\bf Corollary 5.1.} {\em Let $q$ be an odd prime power and $n$ be an odd positive integer. For any odd positive integer $t$ satisfying $1 \leq t \leq n-2$ we have a negacyclic LCD code with the same parameters as the that of the BCH code ${\bf C}_{q, n, t+2, \frac{n-t}{2}}$.}\\

{\bf Proof.} The conclusion follows from Theorem 13 in \cite{LDL}.\\

From a similar argument we have the following result.\\

{\bf Corollary 5.2.} {\em Let $q$ be an odd prime power, $n$ be an even number, and ${\bf C} \subset {\bf F}_q^n$ be a cyclic LCD code with the even length $n$. Suppose that ${\bf C}$ is generated by a factor ${\bf g}(x)$ of $x^n-1$. Then we have a cyclic LCD code ${\bf C}'$ generated  by $g(-x)$.}\\

From Corollary 5.2 several families of constructed LCD BCH codes in \cite{LLDL} can be transformed to new families of LCD BCH codes. For example when $q=3$ and the length $n$ is even, if ${\bf C}_{q, n, \delta, b}$ is an LCD BCH code constructed in \cite{LLDL}, then ${\bf C}_{q, n, \delta, b+\frac{n}{2}}$ is also an LCD BCH code.\\

The above variable transformation can be replaced by $x \longrightarrow \eta \cdot x$, where $\eta \in {\bf F}_q$ is a nonzero element satisfying $\eta^n=\pm1$. \\

{\bf Proposition 5.1.} {\em If ${\bf C} \subset {\bf F}_q^n$ is a cyclic code generated by degree $k$ polynomial ${\bf g}(x)$, and $\eta$ is an element in ${\bf F}_q$ such that $\eta^n=1$. Set ${\bf v}=(1, \eta, \eta^2, \ldots, \eta^{n-1})$. Then ${\bf v} \cdot {\bf C}$ is an equivalent cyclic code generated by ${\bf g}(\eta x)$. If $\eta^n=-1$, ${\bf v} \cdot {\bf C}$ is a negacyclic code generated by ${\bf g}(\eta x)$. The dual code of ${\bf v} \cdot {\bf C}$ is generated by $$\frac{(\eta x)^k {\bf h}(\eta^{-1} x^{-1})}{{\bf h}(0)},$$ where ${\bf h}(x)=\frac{x^n-1}{{\bf g}(x)}$.}\\

{\bf Proof.} If $(c_0, c_1, \ldots, c_{n-2}, c_{n-1}) \in {\bf C}$, then $(c_{n-1}, \eta c_1, \ldots, \eta^{n-1} c_{n-2}) \in {\bf v} \cdot {\bf C}$. we need to check if $(\eta^{n-1} c_{n-1}, c_0, \eta c_1, \ldots, \eta^{n-2} c_{n-2})$ is also in ${\bf v} \cdot {\bf C}$. It is obvious $$\eta \cdot (\eta^{n-1} c_{n-1}, c_0, \eta c_1, \ldots, \eta^{n-2} c_{n-2})=(c_{n-1}, \eta c_0, \ldots, \eta^{n-1} c_{n-2}).$$ The conclusion follows. The other conclusion can be proved similarly.\\

Then the hull-variation problem can be restricted to equivalent cyclic or negacyclic codes. It is interesting to consider the hull-variation problem of cyclic codes with respect to the equivalent transformations in Proposition 5.1.\\

\section{Arbitrary dimension hull MDS codes}

Let ${\bf C}_1$ be a linear $[n, k, d]_q$ code. If its dual (or Hermitian dual) ${\bf C}_1^{\perp}$ is of the form ${\bf x} \cdot {\bf C}_2$, where ${\bf x}=(x_1, \ldots, x_n)$ is a Hamming weight $n$ vector in ${\bf F}_q^n$ and both codes ${\bf C}_1$ and ${\bf C}_2$ are in the same family of algebraic codes, then the dual of ${\bf v} \cdot {\bf C}_1$ is ${\bf v}^{-1} \cdot {\bf x} \cdot {\bf C}_2$, where ${\bf v}=(v_1, \ldots, v_n) \in {\bf F}_q^n$ is a Hamming weight $n$ vector. The problem to determine the dimension of the hull $({\bf v} \cdot {\bf C}_1) \bigcap ({\bf v} \cdot {\bf C}_1)^{\perp}$ is equivalent to determine the dimension of ${\bf C}_1 \bigcap {\bf v}^{-2} \cdot {\bf x} \cdot {\bf C}_2$. When $q$ is an even prime power, $v_1^2, \ldots, v_2^2$ can be arbitrary nonzero elements in ${\bf F}_q$. Then this is equivalent to determine the dimension of ${\bf C}_1 \bigcap {\bf u} \cdot {\bf C}_2$, where ${\bf u}=(u_1, \ldots, u_n)$ is an arbitrary Hamming weight $n$ vector in ${\bf F}_q^n$.\\

\subsection{Generalized Reed-Solomon codes}

The following result was frequently used in \cite{JLLX,JX17,FFLZ20} to determine the dual of  the generalized Reed-Solomon codes, we refer to Theorem 5.3.3, page 176 of \cite{HP}.\\

{\bf Lemma 6.1 (Theorem 5.3.3 in \cite{HP}).} {\em Let $a_1, \ldots, a_n \in {\bf F}_q$ be $n$ distinct elements. We consider the following $(n-1) \times n$ Vandermonde matrix ${\bf V}(a_1, \ldots, a_n)$

$$
\left(
\begin{array}{ccccccccc}
1&1&1&\cdots&\cdots&1\\
a_1&a_2&a_3&\cdots&\cdots&a_n\\
a_1^2&a_2^2&a_3^2&\cdots&\cdots&a_n^2\\
\cdots&\cdots&\cdots&\cdots&\cdots&\cdots\\
a_1^{n-2}&a_2^{n-2}&a_3^{n-2}&\cdots&\cdots&a_n^{n-2}\\
\end{array}
\right)
$$
Then the system of linear equations ${\bf V}(a_1, \ldots, a_n) \cdot {\bf x}^{\tau}={\bf 0}$, where ${\bf x} \in {\bf F}_q^n$, has a nonzero solution ${\bf x}$ such that each coordinate of ${\bf x}$ is not zero.}\\

Let ${\bf F}_q$ be an arbitrary finite field, $P_1,\ldots,P_n$ be $n \leq q$ elements in ${\bf F}_q$. A Reed-Solomon code $RS(n, k)$ is defined by $$RS(n,k)=\{(f(P_1),\ldots,f(P_n)): f \in {\bf F}_q[x],\deg(f) \leq k-1\}.$$ This is a linear $[n,k,n-k+1]_q$ MDS codes from the fact that a degree $\deg(f) \leq k-1$ polynomial has at most $k-1$ roots. From Lemma 6.1 the dual of $RS(n, k)$ is $RS(n, k)^{\perp}={\bf x} \cdot RS(n, n-k)$, where ${\bf x}$ is a solution of the system of linear equation in Lemma 6.1. This Hamming weight $n$ vector ${\bf x}$ can be calculated explicitly, see \cite{JLLX,JX17}. In some papers following the idea of \cite{JLLX,JX17} the explicit calculation of coordinates of ${\bf x}$ is needed, see \cite{LCC,FFLZ20,ZhangFeng}. We show that for generalized Reed-Solomon codes over finite fields of the  even characteristic, this is actually not necessary. Notice that for the generalized Reed-Solomon code ${\bf v} \cdot RS(n, k)$, the Euclid dual is ${\bf v}^{-1} \cdot {\bf x} \cdot RS(n, n-k)$. Then it is equivalent to calculate the dimension of  $RS(n, k) \bigcap {\bf v}^{-2} \cdot {\bf x} \cdot RS(n, n-k)$.\\

When $q$ is an even prime power, if we need to construct a generalized Reed-Solomon code with $h$ dimension Euclid hull, it is equivalent to find a suitable Hamming weight $n$ vector ${\bf u} \in {\bf F}_q^n$ such that $$\dim_{{\bf F}_q}(RS(n, k) \bigcap {\bf u} \cdot RS(n, n-k))=h,$$ where ${\bf u}={\bf v}^{-2} \cdot {\bf x}$ can be an arbitrary Hamming weight $n$ vector in ${\bf F}_q^n$, since each element in ${\bf F}_q$ is a square. Thus for codes over finite fields with the even characteristic, the adjustment of ${\bf v}$ makes ${\bf v}^{-2} \cdot {\bf x}$ arbitrary. In particular this implies that for any even code length $n \leq q$, a self-dual generalized Reed-Solomon code can be constructed immediately. This recovers the classical result in \cite{GG08,Gulliver}. \\

Similarly we take a Hamming weight $n$ vector of the form $${\bf u}=(1, \ldots, 1, \lambda, \ldots, \lambda)$$ with $\max\{k+1, n-k+1\}$ coordinates equal to $1$ and $n-\max\{k+1, n-k+1\}$ coordinates equal to $\lambda \neq 1$. Then $RS(n, k) \bigcap {\bf u} \cdot RS(n, n-k)=\{{\bf 0}\}$, since there are no nonzero polynomials ${\bf g}_1$ and ${\bf g}_2$ with degree $\deg({\bf g}_i) \leq \max\{k-1, n-k-1\}$, $i=1, 2$,  such that they have at least $\max\{k+1, n-k+1\}$ same coordinate but not the same. Then the following result recovers one construction in \cite{J2017} immediately. These kinds of codes have been covered in \cite{Carlet2}.\\

{\bf Proposition 6.1.} {\em Let $q$ be an even prime power. Let $n$ be a positive integer satisfying $n \leq q$ and $k$ be a positive integer satisfying $1 \leq k \leq n-2$. Then an explicit LCD MDS generalized Reed Solomon $[n, k, n-k+1]_q$ code is constructed.}\\

The same method gives the following conclusion of length $n \leq q-1$ arbitrary dimension hull MDS generalized Reed-Solomon codes over ${\bf F}_{2^s}$.\\

{\bf Theorem 6.1.} {\em Let $q$ be an even prime power. Let $n$ and $k$ be two positive integers satisfying $k<n\leq q-1$. Then for any non-negative integer $0 \leq l \leq \min\{k, n-k\}$, we can construct a generalized Reed-Solomon code with the $l$-dimension hull. Therefore generalized Reed-Solomon codes with arbitrary dimension hull of arbitrary length $n \leq q-1$ can be constructed explicitly.}\\

{\bf Proof.} Let $\alpha_1, \ldots, \alpha_n$ be $n \leq q-1$ nonzero distinct elements of ${\bf F}_q$. We can assume $k \leq n-k$. For any $l \leq k$, we take the Hamming weight $n$ vector ${\bf u}=(\alpha_1^{k-l}, \ldots, \alpha_n^{k-l})$. Then the codewords of the code ${\bf u} \cdot RS(n, n-k)$ are the evaluations of polynomials ${\bf f}(x)=a_0x^{k-l}+a_1x^{k-l+1}+\cdots+a_{n-k-1}x^{n-l-1}$, $a_0, \ldots, a_{n-k-1} \in {\bf F}_q$, at $\alpha_1, \ldots, \alpha_n$. Now we calculate the dimension of $\dim(RS(n, k) \bigcap {\bf u} \cdot RS(n, n-k))$. An element in the intersection corresponds to ${\bf g}=g_0+g_1x+\cdots+g_{k-1}x^{k-1}$ and ${\bf f}$ of the above form satisfying that their evaluations at $n$ elements $\alpha_1, \ldots, \alpha_n$ are the same. Since the degrees of these two polynomials are less than or equal to $n-1$. Then ${\bf g}$ and ${\bf f}$ are the same polynomial. The only possible ${\bf f}$ has to be the form $a_0x^{k-l}+a_1x^{k-l+1}+\cdots+a_{l-1}x^{k-1}$, where $a_0, \ldots, a_{l-1} \in {\bf F}_q$. The conclusion is proved.\\

In previous papers \cite{LCC,FFLZ20} arbitrary dimension Euclidean hull $q$-ary generalized Reed-Solomon codes for odd and even prime power $q$ with specific lengths were constructed. Our above simple construction is much stronger than their codes when $q$ is an even prime power and solve this problem for almost all lengths.\\

{\bf Corollary 6.1.} {\em Let ${\bf C} \subset {\bf F}_{2^s}^n$ be a length $n \leq 2^s-1$ Reed-Solomon $[n, k]_{2^s}$ code, then the maximal Euclidean hull dimension $\max_{Ehull}({\bf C})=\min \{k, n-k\}$.}\\

The above result is actually a special case of Theorem 2.2.\\

\subsection{Generalized twisted Reed-Solomon codes}

We consider twisted Reed-Solomon codes introduced in \cite{BPN17,BPR}. For a twisted Reed-Solomon code ${\bf C}_{{\bf \alpha}, {\bf t}, {\bf h}, {\bf \eta}}^{n, k}$ evaluated at a subgroup of the multiplicative group ${\bf F}_q^*$, the dual is equivalent to another twisted Reed-Solomon code ${\bf C}_{{\bf \alpha}, k-{\bf h}, n-k-{\bf t}, -{\bf \eta}}^{n, n-k}$. If $q=2^s$, then for suitable ${\bf h}, {\bf t}$ and ${\bf \eta}$, from our method as above, it is easy to get a self-dual twisted Reed-Solomon code. Some of them are MDS, this recovers some self-dual MDS codes from twisted Reed-Solomon codes constructed in \cite{WHL}.\\

In Subsection 6.2 we construct almost arbitrary dimension hull code from the following simple twisted Reed-Solomon codes over the finite field ${\bf F}_{q}$, $q$ is an even prime power, see \cite{BPN17,BPR}. Let $n$ and $k$ be two positive integers satisfying $n \leq q-1$ and $k \leq n-1$. Let $\alpha_1, \ldots, \alpha_n$ be $n$ distinct elements in the finite field ${\bf F}_q$ such that ${\bf \alpha}=\{\alpha_1, \ldots, \alpha_n\}$ is a multiplicative subgroup of ${\bf F}_q^*$. Let $\eta$ be a nonzero element of ${\bf F}_q$. Set ${\bf g}_0=1+\eta x^k$, ${\bf g}_1=x$, \ldots, ${\bf g}_{k-1}=x^{k-1}$. Let ${\bf P}(\eta, k)$ be the linear span over ${\bf F}_q$ by ${\bf g}_0, \ldots, {\bf g}_{k-1}$. The linear code ${\bf C}_{{\bf \alpha}, \eta, k}$ is the evaluation code of these polynomials in ${\bf P}(\eta, k)$ at the above $n$ distinct elements of the subgroup ${\bf \alpha}$. The dimension of the Schur square of ${\bf C}_{{\bf \alpha}, \eta, k}$ is at least $2k$. Thus this code is not equivalent to a Reed-Solomon code when $2k \leq n$, even when it is MDS, see \cite{BPR}.\\

Set ${\bf h}_0=1$, ${\bf h}_1=x$, \ldots, ${\bf h}_{n-k-1}=x^{n-k-1}-\eta x^{n-1}$. Let ${\bf P}(-\eta, n-k)^{\perp}$ be the linear span over ${\bf F}_q$ by ${\bf h}_0, \ldots, {\bf h}_{n-k-1}$. Let ${\bf C}_{{\bf \alpha}, -\eta, n-k}$ be the linear code which is the evaluation code of these polynomials in ${\bf P}(-\eta, n-k)^{\perp}$ at the above $n$ distinct elements of the subgroup ${\bf \alpha}$.\\

{\bf Proposition 6.2 (or see Theorem 5 in \cite{BPR}).} {\em The dual code of ${\bf C}_{{\bf \alpha}, \eta, k}$ is of the form ${\bf x} \cdot {\bf C}_{{\bf \alpha}, -\eta, n-k}$, where ${\bf x} \in {\bf F}_q^n$ is a Hamming weight $n$ vector as in Lemma 6.1.}\\

{\bf Proof.} First of all ${\bf C}_{{\bf \alpha}, \eta, k} \subset RS(n, k+1)$, then ${\bf x} \cdot RS(n, n-k-2) \subset {\bf C}_{{\bf \alpha}, \eta, k}^{\perp}$. We only need to prove that the products of ${\bf h}_{n-k-1}$ and ${\bf g}_0, \ldots, {\bf g}_{k-1}$ contain no terms $x^{v}$ satisfying $v \geq n-1$. Because ${\bf \alpha}=\{\alpha_1, \alpha_2, \ldots, \alpha_n\}$ is a multiplicative subgroup of ${\bf F}_q^*$, then $x^n=1$ for each element in this subgroup. Hence $(1+\eta x^k)(x^{n-k-1}-\eta x^{n-1})=x^{n-k-1}-\eta^2 x^{k-1}$, and $h_{n-k-1}x^i=x^{n-k-1+i}-\eta x^{i-1}$ for $i=1, \ldots, k-1$, when evaluated at elements of this subgroup ${\bf \alpha}=\{\alpha_1, \ldots, \alpha_n\}$. The conclusion follows immediately.\\

{\bf Proposition 6.3 (or see Theorem 1 of \cite{BPR}).} {\em If $\eta$ is not in the subgroup ${\bf \alpha}$, then ${\bf C}_{{\bf \alpha}, \eta, k}$ is an MDS code.}\\

{\bf Proof} We only need to verify that any nonzero polynomial in ${\bf P}(\eta, k)$ has at most $k-1$ roots in the subgroup ${\bf \alpha}$. The polynomial in ${\bf P}(\eta, k)$ is of the form $a_0 \eta x^k+a_{k-1}x^{k-1}+\cdots+a_1x+a_0$, then for any $k$ roots $\alpha_{i_1} \ldots, \alpha_{i_k}$ in ${\bf \alpha}$, we have $\prod_{j=1}^k \alpha_{i_j}=\pm \frac{a_0}{a_0\eta}$. This is a contradiction.\\

Thus we have some MDS codes which are not equivalent to Reed-Solomon codes, with almost arbitrary dimension hull.\\

{\bf Theorem 6.2.} {\em Let $q$ be an even prime power, $n$ be a positive integer satisfying $n|q-1$ and $k$ be a positive integer satisfying $k \leq \frac{n}{2}$. Then for any non-negative integer $1 \leq l \leq k-2$, there is a generalized twisted Reed-Solomon code as above with the $l$-dimension hull. Therefore generalized twisted Reed-Solomon codes as above with the $l$-dimension hulls, of the length $n|q-1$ can be constructed explicitly.}\\

{\bf Proof.} For integer $l \leq k$, we take the Hamming weight $n$ vector ${\bf u}=(\alpha_1^{k-l}, \ldots, \alpha_n^{k-l})$. Then the codewords of the code ${\bf u} \cdot {\bf C}_{{\bf \alpha}, -\eta, n-k}$ is the evaluations of polynomials ${\bf f}(x)=a_0x^{k-l}+a_1x^{k-l+1}+\cdots+a_{n-k-2}x^{n-l-2}+a_{n-k-1}(x^{n-l-1}-\eta x^{n+k-l-1})$, $a_0, \ldots, a_{n-k-1} \in {\bf F}_q$. Since this polynomial is evaluated at the subgroup ${\bf \alpha}=\{\alpha_1, \ldots, \alpha_n\}$, then ${\bf f}(x)=a_0x^{k-l}+a_1x^{k-l+1}+\cdots+a_{n-k-2}x^{n-l-2}+a_{n-k-1}(x^{n-l-1}-\eta x^{k-l-1})$.\\

Now we calculate the dimension $\dim({\bf C}_{{\bf \alpha}, \eta, k} \bigcap {\bf u} \cdot {\bf C}_{{\alpha}, -\eta, n-k})$. An element in the intersection corresponds to ${\bf b}=b_0(1+\eta x^k)+b_1x+\cdots+b_{k-1}x^{k-1}$ and ${\bf f}$ of the above form satisfying that their evaluations at $n$ elements $\alpha_1, \ldots, \alpha_n$ are the same. Because the degrees of these two polynomials are less than or equal to $n-1$. Then ${\bf b}$ and ${\bf f}$ are the same polynomial. If $b_0$ is not zero, then there is a nonzero constant in ${\bf f}$. This is impossible since $k-l-1 \geq 1$. If $a_{n-k-1} \neq 0$, there is a nonzero term $x^{n-l-1}$ in ${\bf b}$, this is impossible since $k<n-l-1$.
The only possible ${\bf f}$ has to be the form $a_0x^{k-l}+a_1x^{k-l+1}+\cdots+a_{l-1}x^{k-1}$, where $a_0, \ldots, a_{l-1} \in {\bf F}_q$. The conclusion is proved.\\

{\bf Corollary 6.2.} {\em Let ${\bf C} \subset {\bf F}_{2^s}^n$ be a length $n|q-1$ twisted Reed-Solomon code as above. Then the maximal Euclidean hull dimension is $\max_{Ehull}({\bf C}) \geq \min \{k-2, n-k-2\}$.}\\

\section{Application to entanglement-assisted quantum codes}

We recall the CSS construction of entanglement-assisted quantum codes from linear codes with the $h$ dimension hull, see \cite{Brun}.\\

{\bf CSS construction of entanglement-assisted quantum codes.} {\em Let ${\bf C} \subset {\bf F}_{q^2}^n$ be a linear $[n, k, d]_{q^2}$ code with the $h$-dimensional Hermitian hull. Then an EAQEC $[[n, k-h, d, n-k-h]]_q$ code and an EAQEC $[[n, n-k-h, d^{\perp}, k-h]]_q$ code can be constructed. When $d \leq \frac{n+2}{2}$ or $d^{\perp} \leq \frac{n+2}{2}$ and the code is MDS, then the constructed code is an MDS EAQEC code.}\\

We present applications to EAQEC codes from the above CSS construction. The following result is direct from Theorem 3.1, Theorem 4.1 and the CSS construction.\\

{\bf Corollary 7.1.} {\em 1) Let $q$ be a prime power and $h$ be a positive integer satisfying $0 \leq h \leq n$. If there exists a linear self-dual $[2n, n, d]_q$ code, then an EAQEC $[[2n, n-h, d, n-h]]_q$ code can be constructed.\\
2)  Let $q$ be a prime power and $h$ be a positive integer satisfying $0 \leq h \leq n$. If there is a Hermitian self-dual $[2n, n, d]_{q^2}$ code, then we can construct an EAQEC $[[2n, n-h, d, n-h]]_q$ code.\\
3) Let $s$ be a positive integer satisfying $s \geq 2$. If there is an LCD  $[n, k, d]_{2^s}$ code ${\bf C}$ satisfying $d\geq 2$ and $d^{\perp} \geq 2$. Moreover the both shortening codes of ${\bf C}$ and ${\bf C}^{\perp}$ at a coordinate position are LCD. Then we can construct an EAQEC $[[n, k-1, d, n-k-1]]_{2^s}$ code.}\\

From the optimal Hermitian self-dual codes constructed in \cite{Sok2} we give the following tables of EAQEC codes over small fields ${\bf F}_2, {\bf F}_3, {\bf F}_4$ and ${\bf F}_5$ from Corollary 7.1. The dimension upper bounds in the quantum Singleton bound are also listed to show that these EAQEC codes are good. In previous papers \cite{Pellikaan,LCC,GYHZ,CZJL,FFLZ20} on construction of EAQEC codes, there is no EAQEC code over small fields presented. Thus we list these codes for the convenience of the practical quantum error correction.\\

\begin{longtable}{|l||l|l|}
\caption{\label{tab:A-q-5-3} $q^2=4$, $q=2$}\\ \hline
EAQEC parameters&h&Singleton\\ \hline
$[[6, 3-h, 4, 3-h]]_2$&$0 \leq h \leq 3$&$3-h$ \\ \hline
$[[8, 4-h, 4, 4-h]]_2$&$0 \leq h \leq 4$&$6-h$  \\ \hline
$[[10, 5-h, 4, 5-h]]_2$&$0 \leq h \leq 5$&$9-h$  \\ \hline
$[[12, 6-h, 4, 6-h]]_2$&$0 \leq h \leq 6$&$12-h$\\ \hline
$[[14, 7-h, 6, 7-h]]_2$&$0 \leq h \leq 7$&$11-h$\\ \hline
$[[16, 8-h, 6, 8-h]]_2$&$0 \leq h \leq 8$&$14-h$\\ \hline
$[[18, 9-h, 8, 9-h]]_2$&$0 \leq h \leq 9$&$13-h$  \\ \hline
$[[22, 11-h, 8, 11-h]]_2$&$0 \leq h \leq 11$&$19-h$ \\ \hline
$[[24, 12-h, 8, 12-h]]_2$&$0 \leq h \leq 12$&$22-h$\\ \hline
$[[26, 13-h, 8, 13-h]]_2$&$0 \leq h \leq 13$&$25-h$ \\ \hline
$[[28, 14-h, 8, 14-h]]_2$ &$0 \leq h \leq 14$&$28-h$  \\ \hline
$[[30, 15-h, 8, 15-h]]_2$&$0 \leq h \leq 15$&$31-h$\\ \hline
$[[34, 17-h, 10, 17-h]]_2$&$0 \leq h \leq 17$&$33-h$\\ \hline
\end{longtable}

\begin{longtable}{|l||l|l|}
\caption{\label{tab:A-q-5-3} $q^2=9$, $q=3$}\\ \hline
EAQEC parameters&h&Singleton\\ \hline
$[[6, 3-h, 4, 3-h]]_3$&$0 \leq h \leq 3$&$3-h$ \\ \hline
$[[8, 4-h, 4, 4-h]]_3$&$0 \leq h \leq 4$&$6-h$  \\ \hline
$[[10, 5-h, 6, 5-h]]_3$&$0 \leq h \leq 5$&$5-h$  \\ \hline
$[[12, 6-h, 6, 6-h]]_3$&$0 \leq h \leq 6$&$8-h$\\ \hline
$[[14, 7-h, 6, 7-h]]_3$&$0 \leq h \leq 7$&$11-h$\\ \hline
$[[16, 8-h, 7, 8-h]]_3$&$0 \leq h \leq 8$&$12-h$\\ \hline
$[[18, 9-h, 7, 9-h]]_3$&$0 \leq h \leq 9$&$15-h$  \\ \hline
$[[20, 10-h, 8, 10-h]]_3$&$0 \leq h \leq 10$&$16-h$\\ \hline
$[[22, 11-h, 8, 11-h]]_3$&$0 \leq h \leq 11$&$19-h$ \\ \hline
$[[24, 12-h, 9, 12-h]]_3$&$0 \leq h \leq 12$&$20-h$\\ \hline
$[[26, 13-h, 9, 13-h]]_3$&$0 \leq h \leq 13$&$23-h$ \\ \hline
$[[28, 14-h, 10, 14-h]]_3$ &$0 \leq h \leq 14$&$22-h$  \\ \hline
\end{longtable}

\begin{longtable}{|l||l|l|}
\caption{\label{tab:A-q-5-3} $q^2=16$, $q=4$}\\ \hline
EAQEC parameters&h&Singleton\\ \hline
$[[6, 3-h, 4, 3-h]]_4$&$0 \leq h \leq 3$&$3-h$ \\ \hline
$[[8, 4-h, 4, 4-h]]_4$&$0 \leq h \leq 4$&$6-h$  \\ \hline
$[[10, 5-h, 6, 5-h]]_4$&$0 \leq h \leq 5$&$5-h$  \\ \hline
$[[12, 6-h, 6, 6-h]]_4$&$0 \leq h \leq 6$&$8-h$\\ \hline
$[[14, 7-h, 7, 7-h]]_4$&$0 \leq h \leq 7$&$9-h$\\ \hline
$[[16, 8-h, 7, 8-h]]_4$&$0 \leq h \leq 8$&$12-h$\\ \hline
$[[18, 9-h, 8, 9-h]]_4$&$0 \leq h \leq 9$&$13-h$  \\ \hline
$[[20, 10-h, 8, 10-h]]_4$&$0 \leq h \leq 10$&$16-h$\\ \hline
$[[22, 11-h, 9, 11-h]]_4$&$0 \leq h \leq 11$&$17-h$ \\ \hline
$[[24, 12-h, 9, 12-h]]_4$&$0 \leq h \leq 12$&$20-h$\\ \hline
$[[26, 13-h, 10, 13-h]]_4$&$0 \leq h \leq 13$&$21-h$ \\ \hline
$[[28, 14-h, 11, 14-h]]_4$ &$0 \leq h \leq 14$&$22-h$  \\ \hline
\end{longtable}

\begin{longtable}{|l||l|l|}
\caption{\label{tab:A-q-5-3} $q^2=25$, $q=5$}\\ \hline
EAQEC parameters&h&Singleton\\ \hline
$[[6, 3-h, 4, 3-h]]_5$&$0 \leq h \leq 3$&$3-h$ \\ \hline
$[[8, 4-h, 5, 4-h]]_5$&$0 \leq h \leq 4$&$4-h$  \\ \hline
$[[10, 5-h, 6, 5-h]]_5$&$0 \leq h \leq 5$&$5-h$  \\ \hline
$[[12, 6-h, 6, 6-h]]_5$&$0 \leq h \leq 6$&$8-h$\\ \hline
$[[14, 7-h, 7, 7-h]]_5$&$0 \leq h \leq 7$&$9-h$\\ \hline
$[[18, 9-h, 8, 9-h]]_5$&$0 \leq h \leq 9$&$13-h$  \\ \hline
$[[22, 11-h, 9, 11-h]]_5$&$0 \leq h \leq 11$&$17-h$ \\ \hline
$[[24, 12-h, 10, 12-h]]_5$&$0 \leq h \leq 12$&$18-h$\\ \hline
$[[26, 13-h, 10, 13-h]]_5$&$0 \leq h \leq 13$&$21-h$ \\ \hline
$[[28, 14-h, 11, 14-h]]_5$ &$0 \leq h \leq 14$&$22-h$  \\ \hline
\end{longtable}

\begin{longtable}{|l||l|l|}
\caption{\label{tab:A-q-5-3} $q^2=49, 64, 81, 121, 169$, $q=7, 8, 9, 11, 13$}\\ \hline
EAQEC parameters&h&Singleton\\ \hline
$[[10, 5-h, 6, 5-h]]_q$&$0 \leq h \leq 5$&$5-h$  \\ \hline
$[[12, 6-h, 7, 6-h]]_q$&$0 \leq h \leq 6$&$6-h$\\ \hline
$[[14, 7-h, 7, 7-h]]_q$&$0 \leq h \leq 7$&$9-h$\\ \hline
$[[16, 9-h, 8, 9-h]]_q$&$0 \leq h \leq 8$&$10-h$  \\ \hline
$[[18, 9-h, 9, 9-h]]_q$&$0 \leq h \leq 9$&$11-h$ \\ \hline
\end{longtable}

\begin{longtable}{|l||l|l|}
\caption{\label{tab:A-q-5-3} $q^2=256, 289, 361$, $q=16, 17, 19$}\\ \hline
EAQEC parameters&h&Singleton\\ \hline
$[[10, 5-h, 6, 5-h]]_q$&$0 \leq h \leq 5$&$5-h$  \\ \hline
$[[12, 6-h, 7, 6-h]]_q$&$0 \leq h \leq 6$&$6-h$\\ \hline
$[[14, 7-h, 8, 7-h]]_q$&$0 \leq h \leq 7$&$7-h$\\ \hline
$[[16, 9-h, 8, 9-h]]_q$&$0 \leq h \leq 8$&$10-h$  \\ \hline
$[[18, 9-h, 9, 9-h]]_q$&$0 \leq h \leq 9$&$11-h$ \\ \hline
\end{longtable}

There are many constructions of MDS EAQEC code and almost MDS EAQEC codes in \cite{Pellikaan,LCC,GYHZ,CZJL}. We have the following result.\\

{\bf Corollary 7.2.} {\em Let $q$ be an even prime power satisfying $q \geq 4$, $n$, $k\leq \frac{n}{2}$ and $h$ be three positive integers satisfying  $n \leq q-1$, $0\leq h \leq k$. Then an MDS EAQEC $[[n, n-k-h, k+1, k-h]]_q$ code can be constructed.}\\

{\bf Proof.} The result follows from Theorem 6.1 immediately.\\

Notice that from Theorem 3.1 for all lengths $n \leq q-1$, the parameter $c$ can be arbitrary in the range $0\leq c \leq k$. The only restriction on $k$ is $k \leq n-1$. Hence these MDS EAQEC codes in Corollary 7.2 are new comparing to the constructions in \cite{LCC,FFLZ20,GYHZ,MCao}. The parameters of EAQEC codes constructed in Corollary 7.2 are flexible than previous codes constructed in \cite{LCC,FFLZ20,GYHZ,MCao}. Moreover these MDS codes can be constructed from almost arbitrary dimension hull generalized twisted Reed-Solomon codes in Theorem 6.2 when the length $n$ satisfies $n|q-1$.\\

From elliptic curve codes defined over the finite field ${\bf F}_{2^s}$, for any even length $4 \leq n \leq 2^s+\lfloor 2^{\frac{s}{2}+1} \rfloor-2$, self-dual near MDS (then  almost MDS) codes were constructed in \cite{JK19}. From Theorem 3.1 we have the following new family of almost MDS EAQEC codes. Actually arbitrary dimension hull near MDS codes can be obtained from these self-dual near MDS codes in \cite{JK19} when $q$ is an even prime power, from Theorem 3.1 of this paper.\\

{\bf Corollary 7.3.} {\em Let $n$ be an even positive integer satisfying $4 \leq n \leq 2^s+\lfloor 2^{\frac{s}{2}+1} \rfloor-2$. We can construct an almost MDS EAQEC $[[n, \frac{n}{2}-h, \geq \frac{n}{2}, \frac{n}{2}-h]]_{2^s}$ code for any nonnegative integer $h$ satisfying $0 \leq h \leq \frac{n}{2}$.}\\

Comparing with the almost MDS EAQEC codes constructed in \cite{Pellikaan} from elliptic curves, these codes are new.\\

\section{Conclusion}

We proposed to study the hull-variation problem and introduced the maximal Euclidean (or Hermitian) hull dimension as an invariant of a linear code with respect to equivalent transformations. Some basic properties of maximal Euclidean (Hermitian) hull dimensions were given. It was proved that a length $n$ (Euclidean or Hermitian) self-dual code is equivalent to a linear $h$-dimension hull code for all possible $h$ satisfying $0\leq h\leq \frac{n}{2}$. We proved that a linear LCD code with the minimum Hamming distance $d \geq 2$ and the dual distance $d^{\perp} \geq 2$ has an equivalent linear one-dimension hull code under a weak condition, where $q$ is an even prime power satisfying $q\geq 4$. Then it follows that for any fixed nonnegative integer $h$, there is an asymptotically good sequence of linear $h$-dimension hull codes over ${\bf F}_q$ better than the Gilbert-Varshamov bound. We also showed that this method gives many new arbitrary dimension hull generalized Reed-Solomon codes over ${\bf F}_{2^s}$ with arbitrary length $n \leq q-1$. Arbitrary dimension hull generalized twisted Reed-Solomon codes over ${\bf F}_{2^s}$ with the length $n|q-1$ were also constructed. Applying our results to entanglement-assisted quantum error correction codes some new MDS and new almost MDS EAQEC codes were constructed. Many EAQEC codes over small fields were constructed from optimal Hermitian self-dual codes. The hull-increasing variation problem of general linear codes is still interesting to understand further.\\

{\bf Acknowledgement.} The author thanks Professor Chunming Tang, Professor Chengju Li and Dr. Gaojun Luo for helpful discussion on the hull-variation problem. The author thanks the Associate Editor and four reviewers sincerely for their helpful comments that improved the presentation of this paper.

\end{document}